\documentclass[aps,pra,11pt,tightenlines,eqsecnum,longbibliography,notitlepage,amsmath,onecolumn,superscriptaddress]{revtex4-1}
\usepackage[dvipsnames]{xcolor}
\usepackage[colorlinks=true,bookmarks=false,linkcolor=NavyBlue,urlcolor=NavyBlue,citecolor=NavyBlue,breaklinks]{hyperref}
\usepackage{amsmath}
\usepackage{amsfonts}
\usepackage{tabularx}
\usepackage{graphicx}
\usepackage{times}
\usepackage{braket}

\newcommand{\parti}[2]{\frac{\partial #1}{\partial #2}}

\newcommand{\bs}[1]{\boldsymbol{#1}}
\newcommand{\abs}[1]{\left|#1\right|}
\newcommand{\bk}[1]{\left(#1\right)}
\newcommand{\Bk}[1]{\left[#1\right]}

\newcommand{\expect}{\mathbb E}

\begin{document}
\title{Subdiffraction incoherent optical imaging via spatial-mode
  demultiplexing}

\author{Mankei Tsang}
\email{mankei@nus.edu.sg}
\affiliation{Department of Electrical and Computer Engineering,
  National University of Singapore, 4 Engineering Drive 3, Singapore
  117583}

\affiliation{Department of Physics, National University of Singapore,
  2 Science Drive 3, Singapore 117551}

\date{\today}

\begin{abstract}
  I propose a spatial-mode demultiplexing (SPADE) measurement scheme
  for the far-field imaging of spatially incoherent optical
  sources. For any object too small to be resolved by direct imaging
  under the diffraction limit, I show that SPADE can estimate its
  second or higher moments much more precisely than direct imaging can
  fundamentally do in the presence of photon shot noise. I also prove
  that SPADE can approach the optimal precision allowed by quantum
  mechanics in estimating the location and scale parameters of a
  subdiffraction object. Realizable with far-field linear optics and
  photon counting, SPADE is expected to find applications in both
  fluorescence microscopy and astronomy.
\end{abstract}

\maketitle

\section{Introduction}
Recent research, initiated by our group
\cite{tnl,sliver,tnl2,nair_tsang16,tsang16,ant,lu16}, has shown that
far-field linear optical methods can significantly improve the
resolution of two equally bright incoherent optical point sources with
sub-Rayleigh separations
\cite{tnl,sliver,tnl2,nair_tsang16,tsang16,ant,lu16,lupo,rehacek16,krovi16,kerviche17,tang16,yang16,tham16,paur16},
overcoming previously established statistical limits
\cite{bettens,vanaert,ram,acuna}.  The rapid experimental
demonstrations \cite{tang16,yang16,tham16,paur16} have heightened the
promise of our approach.  An open problem, of fundamental interest in
optics and monumental importance to astronomy and fluorescence
microscopy, is whether these results can be generalized for an
arbitrary distribution of incoherent sources.  Here I take a major
step towards solving the problem by proposing a generalized
spatial-mode demultiplexing (SPADE) measurement scheme and proving its
superiority over direct imaging via a statistical analysis.

The use of coherent optical processing to improve the lateral
resolution of incoherent imaging has thus far received only modest
attention, as prior proposals
\cite{walker93,tamburini,sandeau06,wicker07,wicker09,weigel,yang16}
either did not demonstrate any substantial improvement or neglected
the important effect of noise.  Using quantum optics and parameter
estimation theory, here I show that, for any object too small to be
resolved by diffraction-limited direct imaging, SPADE can estimate its
second or higher moments much more precisely than direct imaging can
fundamentally do in the presence of photon shot noise. Moreover, I
prove that SPADE can approach the optimal precision allowed by quantum
mechanics in estimating the location and scale parameters of a
subdiffraction object.  Given the usefulness of moments in identifying
the size and shape of an object \cite{prokop}, the proposed scheme,
realizable with far-field linear optics and photon counting, should
provide a major boost to incoherent imaging applications that are
limited by diffraction and photon shot noise, including not only
fluorescence microscopy \cite{pawley,deschout,mortensen10,chao16} and
space-based telescopes \cite{huber} but also modern ground-based
telescopes \cite{christou16,sanders13,bernstein14,davies16}.

This paper is organized as follows. Section~\ref{sec_background}
introduces the background theory of quantum optics and parameter
estimation for incoherent imaging. Section~\ref{sec_spade} describes
the SPADE scheme for general imaging.  Section~\ref{sec_stat} presents
the most important results of this paper, namely, a comparison between
the statistical performances of direct imaging and SPADE in the
subdiffraction regime, showing the possibility of giant precision
enhancements for moment estimation, while Appendix~\ref{app_nuisance}
justifies an approximation made in Sec.~\ref{sec_stat} in more detail.
Section~\ref{sec_numerical} presents a numerical example to illustrate
the theory, comparing the errors in estimating the first and second
moments of subdiffraction objects using direct imaging and
SPADE. Section~\ref{sec_limits} proves that SPADE is close to the
quantum precision limits to location and scale estimation in the
subdiffraction regime.  Section~\ref{sec_discuss} discusses other
practical and open issues.

\section{Background formalism}
\label{sec_background}
\subsection{Quantum optics}
I begin with the quantum formalism established in Ref.~\cite{tnl} to
ensure correct physics. The quantum state of thermal light with $M$
temporal modes and a bandwidth much smaller than the center frequency
can be written as $\rho^{\otimes M}$, where
\begin{align}
\rho = (1-\epsilon)\rho_0 +\epsilon \rho_1 +O(\epsilon^2),
\end{align}
$\epsilon$ is the average photon number per mode assumed to be $\ll 1$
\cite{zmuidzinas03,goodman_stat},
$\rho_0 = \ket{\textrm{vac}}\bra{\textrm{vac}}$ is the vacuum state,
$\rho_1$ is the one-photon state with a density matrix equal to the
mutual coherence function, and $O(\epsilon^2)$ denotes second-order
terms, which are neglected hereafter. It is standard to assume that
the fields from incoherent objects, such as stellar or fluorescent
emitters, are spatially uncorrelated at the source
\cite{goodman_stat}.  In a diffraction-limited imaging system, the
fields then propagate as waves; the Van Cittert-Zernike theorem is the
most venerable consequence \cite{goodman_stat}. At the image plane of
a conventional lens-based two-dimensional imaging system in the
paraxial regime \cite{goodman,goodman_stat}, this implies
\begin{align}
\rho_1 &= \int d^2\bs R  F(\bs R)\ket{\psi_{\bs R}}\bra{\psi_{\bs R}},
\label{rho1}
&
\ket{\psi_{\bs R}} &= \int d^2\bs r \psi(\bs r-\bs R)\ket{\bs r},
\end{align}
where $\bs R = (X,Y)$ is the object-plane position, the notation
$(u_1,u_2,\dots)$ denotes a column vector, $F(\bs R)$ is the source
intensity distribution with normalization
$\int d^2\bs R F(\bs R) = 1$,
$\ket{\bs r} = a^\dagger(\bs r)\ket{\textrm{vac}}$ is a one-photon
position eigenket on the image plane at position $\bs r = (x,y)$ with
$[a(\bs r),a^\dagger(\bs r')] = \delta^2(\bs r-\bs r')$
\cite{shapiro09}, and $\psi(\bs r)$ is the field point-spread function
(PSF) of the imaging system.  Without loss of generality, the
image-plane position vector $\bs r$ has been scaled with respect to
the magnification to follow the same scale as $\bs R$
\cite{goodman}. For convenience, I also normalize the position vectors
with respect to the width of the PSF to make them dimensionless.

Consider the processing and measurement of the image-plane field by
linear optics and photon counting. The counting distribution for each
$\rho$ can be expressed as
$\bra{n_0,n_1,\dots}\rho \ket{n_0,n_1,\dots}$, where
$\ket{n_0,n_1,\dots} = (\prod_{j=0}^\infty b_j^{\dagger
  n_j}/\sqrt{n_j!})  \ket{\textrm{vac}}$,
$b_j \equiv \int d^2\bs r \phi_j^*(\bs r) a(\bs r)$,
$\phi_j(\bs r)$ is the optical mode function that is projected to
the $j$th output, and
$[b_j,b_k^\dagger] = \int d^2\bs r \phi_j^*(\bs r)\phi_k(\bs r)
= \delta_{jk}$. With the negligence of multiphoton coincidences, the
relevant projections are $\{\ket{\textrm{vac}},\ket{\phi_j}\}$,
with
$\ket{\phi_j} \equiv \ket{0,\dots,n_j = 1,\dots, 0} = b_j^\dagger
\ket{\textrm{vac}} = \int d^2\bs r \phi_j(\bs r)\ket{\bs r}$.  The
zero-photon probability becomes $1-\epsilon$ and the probability of
one photon being detected in the $j$th mode becomes $\epsilon p(j)$,
where
\begin{align}
p(j) \equiv  \bra{\phi_j}\rho_1\ket{\phi_j}
=\int d^2\bs R F(\bs R)\abs{\braket{\phi_j|\psi_{\bs R}}}^2
\label{pj}
\end{align}
is the one-photon distribution.  A generalization of the measurement
model using the concept of positive operator-valued measures is
possible \cite{tnl,tnl2} but not needed here.

For example, direct imaging can be idealized as a measurement of the
position of each photon, leading to an expected image given by
\begin{align}
f(\bs r) &\equiv \bra{\bs r}\rho_1\ket{\bs r} = 
\int d^2\bs R  F(\bs R)\abs{\psi(\bs r-\bs R)}^2,
\label{lambda}
\end{align}
which is a basic result in statistical optics
\cite{goodman_stat,goodman}.  While Eq.~(\ref{lambda}) suggests that,
similar to the coherent-imaging formalism, the PSF acts as a low-pass
filter in the spatial frequency domain \cite{goodman}, the effect of
more general optical processing according to Eq.~(\ref{pj}) is more
subtle and offers surprising advantages, as demonstrated by recent
work
\cite{tnl,ant,sliver,nair_tsang16,tnl2,tsang16,lu16,lupo,rehacek16,tang16,yang16,tham16,paur16,krovi16,kerviche17}
and elaborated in this paper.

Over $M$ temporal modes, the probability distribution of photon
numbers $m = (m_0,m_1,\dots)$ detected in the respective optical modes
becomes
\begin{align}
P(m) &= \sum_{L} \mathcal M(m|L) \mathcal B(L),
\label{Pm}
\end{align}
where $\mathcal B(L)$ is the binomial distribution for detecting $L$
photons over $M$ trials with single-trial success probability
$\epsilon$ and
$\mathcal M (m|L) = \delta_{L,\sum_jm_j} L! \prod_j [p(j)]^{m_j}/m_j!$
is the multinomial distribution of $m$ given $L$ total photons
\cite{wasserman}. The average photon number in all modes becomes
$N \equiv M\epsilon$.  Taking the limit of $\epsilon \to 0$ while
holding $N$ constant, $\mathcal B(L)$ becomes Poisson with mean $N$,
and $P(m) \to \exp(-N)\prod_j [Np(j)]^{m_j}/m_j!$, which is the widely
used Poisson model of photon counting for incoherent sources at
optical frequencies
\cite{tnl2,ram,deschout,pawley,mortensen10,zmuidzinas03,huber,chao16}.

\subsection{Parameter estimation}
The central goal of imaging is to infer unknown properties of the
source distribution $F(\bs R)$ from the measurement outcome $m$. Here
I frame it as a parameter estimation problem, defining
$\theta = (\theta_1,\theta_2,\dots)$ as a column vector of unknown
parameters and assuming the source distribution $F(\bs R|\theta)$ to
be a function of $\theta$.  Denote an estimator as $\check\theta(m)$
and its error covariance matrix as
$\Sigma_{\mu\nu}(\theta) = \sum_m
P(m|\theta)[\check\theta_\mu(m)-\theta_\mu]
[\check\theta_\nu(m)-\theta_\nu]$.  For any unbiased estimator
($\sum_m \check\theta(m) P(m|\theta) = \theta$), the Cram\'er-Rao
bound (CRB) is given by \cite{wasserman,vantrees}
\begin{align}
  \Sigma(\theta) &\ge  \textrm{CRB}(\theta) \equiv J^{-1}(\theta),
\end{align}
where $J(\theta)$ is the Fisher information matrix and the matrix
inequality implies that $\Sigma-J^{-1}$ is positive-semidefinite, or
equivalently $u^\top(\Sigma-J^{-1})u \ge 0$ for any real vector
$u$. Assuming the model given by Eq.~(\ref{Pm}) and a known $N$, it
can be shown \cite{tnl2} that
\begin{align}
J_{\mu\nu}(\theta) &= N \sum_j \frac{1}{p(j|\theta)} \parti{p(j|\theta)}{\theta_\mu}
\parti{p(j|\theta)}{\theta_\nu},
\label{fisher}
\end{align}
which is a well known expression
\cite{bettens,vanaert,ram,zmuidzinas03,deschout,chao16}.  For example,
the direct-imaging information, given Eq.~(\ref{lambda}) and the limit
$p(j|\theta) \to d^2\bs r f(\bs r|\theta)$, is
\begin{align}
J_{\mu\nu}^{(\textrm{direct})}(\theta) &= 
N \int d^2\bs r \frac{1}{f(\bs r|\theta)} 
\parti{f(\bs r|\theta)}{\theta_\mu}\parti{f(\bs r|\theta)}{\theta_\nu}.
\label{Jdirect}
\end{align}
For large $N$, the maximum-likelihood estimator is asymptotically
normal with mean $\theta$ and covariance
$\Sigma(\theta) = J^{-1}(\theta)$, even though it may be biased for
finite $N$ \cite{wasserman,vantrees}. Bayesian and minimax
generalizations of the CRB for any biased or unbiased estimator are
possible \cite{tsang16,vantrees} but not considered here as they offer
qualitatively similar conclusions.  The Fisher information is nowadays
regarded as the standard precision measure in incoherent imaging
research, especially in fluorescence microscopy
\cite{deschout,ram,mortensen10,chao16}, where photon shot noise is the
dominant noise source and a proper statistical analysis is essential.

Apart from the CRB, another useful property of the Fisher information
is the data-processing inequality \cite{zamir,hayashi}, which mandates
that, once the measurement is made, no further processing of the data
can increase the information.  For example, direct imaging with large
pixels can be modeled as integrations of photon counts over groups of
infinitesimally small pixels, so the information can never exceed
Eq.~(\ref{Jdirect}). More generally, the data-processing inequality
rules out the possibility of improving the information using any
processing that applies to the direct-imaging intensity, such as the
proposal by Walker \textit{et al.}\ for incoherent imaging in
Ref.~\cite{walker93}, even if the processing is done with
optics. Hence, as argued by Tham \textit{et al.}~\cite{tham16},
coherent processing that is sensitive to the phase of the field is the
only way to improve upon Eq.~(\ref{Jdirect}). The information for any
coherent processing and measurement is in turn limited by quantum
upper bounds in terms of $\rho_1$
\cite{helstrom,holevo11,hayashi05,hayashi,tnl,tnl2,ant}.

\section{Spatial-mode demultiplexing (SPADE)}
\label{sec_spade}
SPADE is a technique previously proposed for the purpose of
superresolving the separation between two incoherent point sources
\cite{tnl,ant,lu16,tham16,yang16,paur16,rehacek16}. I now ask how
SPADE can be generalized for the imaging of an arbitrary source
distribution.  Consider the transverse-electromagnetic (TEM) basis
$\{\ket{\bs q};\bs q = (q_x,q_y)\in\mathbb N^2\}$ \cite{yariv},
where
\begin{align}
\ket{\bs q} &= 
\int d^2\bs r 
\phi_{\bs q}(\bs r)
\ket{\bs r},
\\
\phi_{\bs q}(\bs r) &\equiv 
\frac{\operatorname{He}_{q_x}(x)\operatorname{He}_{q_y}(y)}{\sqrt{2\pi q_x!q_y!}}
\exp\bk{-\frac{x^2+y^2}{4}},
\label{TEM}
\end{align}
and $\operatorname{He}_q$ is the Hermite polynomial
\cite{NIST:DLMF,Olver:2010:NHMF}. Assuming a Gaussian PSF given by
$\psi(\bs r) = \phi_{00}(\bs r)$, which is a common assumption in
fluorescence microscopy \cite{deschout,chao16}, $\ket{\psi_{\bs R}}$
is a coherent state \cite{mandel}, and the one-photon density matrix
in the TEM basis becomes
\begin{align}
g(\bs q,\bs q'|\theta) &\equiv \bra{\bs q}\rho_1(\theta)\ket{\bs q'}
\\
&=  C(\bs q,\bs q')
\int d^2\bs R 
 F(\bs R|\theta)e^{-(X^2+Y^2)/4}
X^{q_x + q_x'}Y^{q_y+q_y'},
\label{g}
\end{align}
where
\begin{align}
C(\bs q,\bs q') &\equiv \frac{1}{2^{|\bs q+\bs q'|_1}\sqrt{\bs q!\bs q'!}},
&
|\bs q|_1 &\equiv q_x + q_y,
&
\bs q! &\equiv q_x! q_y!.
\label{C}
\end{align}

To investigate the imaging capability of SPADE measurements, define
\begin{align}
\Theta_{\bs\mu}(\theta) &\equiv 
\int d^2\bs R 
 F(\bs R|\theta)e^{-(X^2+Y^2)/4}
X^{\mu_X}Y^{\mu_Y},
\label{Theta}
\end{align}
with $\bs\mu = (\mu_X,\mu_Y)$, leading to a linear
parameterization of $g$ given by
\begin{align}
g(\bs q,\bs q'|\theta) &= C(\bs q,\bs q')
\Theta_{\bs q+\bs q'}.
\end{align}
Notice that each $\Theta_{\bs\mu}$ is a moment of the source
distribution filtered by a Gaussian. In particular, if the object is
much smaller than the PSF width, the Gaussian can be neglected, and
$\Theta_{\bs\mu}$ becomes a moment of the source distribution itself.
This subdiffraction regime is of central interest to superresolution
imaging and, as shown in Sec.~\ref{sec_stat}, also a regime in which
direct imaging performs relatively poorly. Since a distribution is
uniquely determined by its moments \cite{bertero89},
$F(\bs R|\theta)\exp[-(X^2+Y^2)/4]$ and therefore $F(\bs R|\theta)$
can be reconstructed given the moments, at least in principle.  Note
also that the object-moment order $\bs\mu$ is nontrivially related to
the order of the matrix element via $\bs\mu = \bs q + \bs q'$, which
is a peculiar feature of incoherent imaging.

A measurement in the TEM basis yields
\begin{align}
p^{(\textrm{TEM})}(\bs q|\theta) &= C(\bs q,\bs q)\Theta_{2\bs q},
\label{pTEM}
\end{align}
which is sensitive only to moments with even $\mu_X$ and $\mu_Y$, as
also recognized by Yang \textit{et al.}\ in Ref.~\cite{yang16}.  This
measurement is realized by demultiplexing the image-plane optical
field in terms of the TEM basis via linear optics before photon
counting for each mode and can be implemented by many methods, most
commonly found in optical communications
\cite{tnl,ant,paur16,armstrong,miller13,li14}.  To access the
other moments, consider interferometry between two TEM modes that
implements the projections
\begin{align}
\ket{+} &\equiv \frac{1}{\sqrt{2}}\bk{\ket{\bs q}+\ket{\bs q'}},
&
\ket{-} &\equiv \frac{1}{\sqrt{2}}\bk{\ket{\bs q}-\ket{\bs q'}}.
\end{align}
This two-channel interferometric TEM (iTEM) measurement leads to
\begin{align}
p^{(\bs q,\bs q')}(+|\theta) &= 
\beta(\bs q,\bs q')+C(\bs q,\bs q')\Theta_{\bs q+\bs q'},
\nonumber\\
p^{(\bs q,\bs q')}(-|\theta) &= 
\beta(\bs q,\bs q')-C(\bs q,\bs q')\Theta_{\bs q+\bs q'},
\label{piTEM} \\
\beta(\bs q,\bs q') &\equiv \frac{1}{2}\Bk{C(\bs q,\bs q)\Theta_{2\bs q}+
C(\bs q',\bs q')\Theta_{2\bs q'}}.
\label{beta}
\end{align}
The dependence on $\Theta_{\bs q+\bs q'}$ is the main interest here,
as it allows one to access any moment parameter.

\begin{figure}[htbp!]
\centerline{\includegraphics[width=0.6\textwidth]{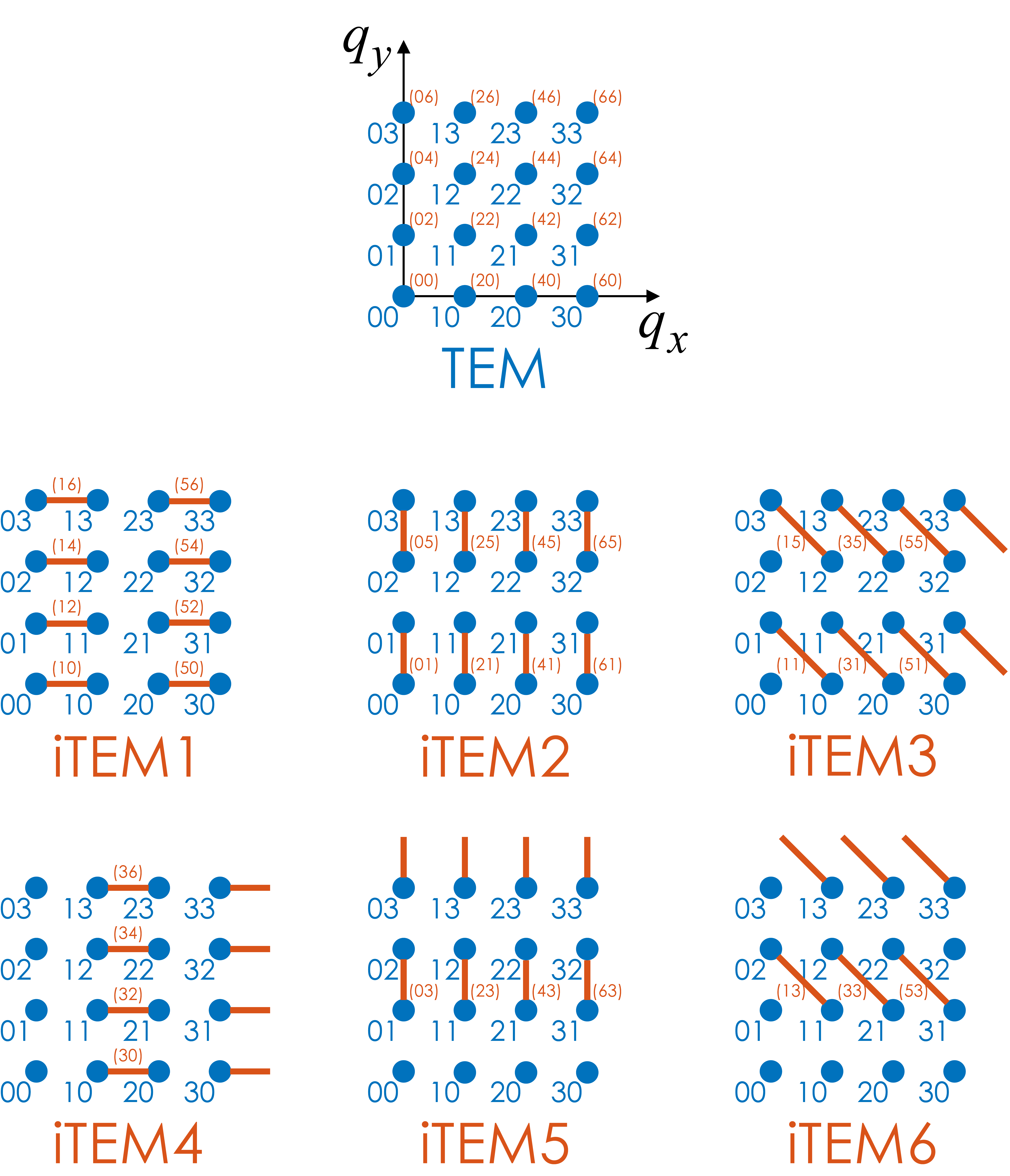}}
\caption{\label{iTEM}Each dot corresponds to a TEM mode in the
  $(q_x,q_y)$ space, and each line connecting two dots denotes an
  interferometer between two modes in an iTEM scheme.  The bracketed
  numbers are the orders $(\mu_X,\mu_Y)$ of the moment parameters to
  which the projections are sensitive. The unconnected dots in some of
  the iTEM schemes denote the rest of the modes in a complete basis,
  which can be measured simultaneously to provide extra information. }
\end{figure}

For multiparameter estimation and general imaging, multiple TEM and
iTEM measurements are needed. To be specific, Table~\ref{projections}
lists a set of schemes that can be used together to estimate all the
moment parameters, while Fig.~\ref{iTEM} shows a graphical
representation of the schemes in the $(q_x,q_y)$ space.  Neighboring
modes are used in the proposed iTEM schemes because they maximize the
Fisher information, as shown later in Sec.~\ref{sec_stat}.  The bases
in different schemes are incompatible with one another, so the photons
have to be rationed among the 7 schemes, by applying the different
schemes sequentially through reprogrammable interferometers or
spatial-light modulators \cite{paur16,armstrong,miller13,li14} for
example.

\begin{table}[htbp!]
\centerline{
\begin{tabularx}{\columnwidth}{|l|l|X|X|X|X|}
\hline 
Scheme & Projections & $q_x$ & $q_y$ 
& $\mu_X$ & $\mu_Y$ \\
\hline
TEM & $\ket{\bs q}$ & $\mathbb N$ & $\mathbb N$ & even & even \\
iTEM1 & $[\ket{\bs q}\pm \ket{\bs q+(1,0)}]/\sqrt{2}$ 
& even & $\mathbb N$ & $1,5,\dots$ & even \\
iTEM2 & $[\ket{\bs q}\pm \ket{\bs q+(0,1)}]/\sqrt{2}$ 
 & $\mathbb N$ & even &  even & $1,5,\dots$\\
iTEM3 & $[\ket{\bs q}\pm \ket{\bs q+(1,-1)}]/\sqrt{2}$ 
 & $\mathbb N$ & odd  &  odd & $1,5,\dots$\\
iTEM4 & $[\ket{\bs q}\pm \ket{\bs q+(1,0)}]/\sqrt{2}$ 
 & odd & $\mathbb N$ & $3,7,\dots$ & even\\
iTEM5 & $[\ket{\bs q}\pm \ket{\bs q+(0,1)}]/\sqrt{2}$ 
 & $\mathbb N$ & odd & even & $3,7,\dots$\\
iTEM6 & $[\ket{\bs q}\pm \ket{\bs q+(1,-1)}]/\sqrt{2}$ 
 & $\mathbb N$ & even & odd & $3,7,\dots$\\
\hline
\end{tabularx}
}
\caption{A list of measurement schemes, their projections,
  and the orders $\bs\mu = (\mu_X,\mu_Y)$ of the moment parameters
  $\Theta_{\bs\mu}$ to which they are sensitive.}
\label{projections}
\end{table}

\section{Statistical analysis}
\label{sec_stat}
\subsection{Direct imaging}
Although the proposed SPADE method can in principle perform general
imaging, its complexity would not be justifiable if it could not offer
any significant advantage over direct imaging. To compare their
statistical performances, consider first direct imaging with a
Gaussian PSF.  Expanding $|\psi(\bs r-\bs R)|^2$ in a Taylor series, I
obtain
\begin{align}
f(\bs r|\theta) &= \abs{\phi_{00}(\bs r)}^2
\Bk{1 +\sum_{\bs\mu} D_{\bs\mu}(\bs r) \theta_{\bs\mu}},
\label{lambda_linear}
\\
D_{\bs\mu}(\bs r) &\equiv \frac{\operatorname{He}_{\mu_X}(x) 
\operatorname{He}_{\mu_Y}(y)}{\bs\mu!},
\end{align}
in terms of the moment parameters defined as
\begin{align}
\theta_{\bs\mu} \equiv \int d^2\bs R  F(\bs R|\theta)
X^{\mu_X} Y^{\mu_Y}.
\end{align}
In terms of this parameterization, the Fisher information becomes
\begin{align}
J_{\bs\mu\bs\nu}^{(\textrm{direct})}
&= N \int d^2\bs r \abs{\phi_{00}(\bs r)}^2
\frac{D_{\bs\mu}(\bs r)D_{\bs\nu}(\bs r)}
{1 + \sum_{\bs\eta} D_{\bs\eta}(\bs r) \theta_{\bs\eta}}.
\label{Jdirect2}
\end{align}
Assume now that the support of the source distribution is centered at
the origin and has a maximum width $\Delta$ much smaller than the PSF
width. Since the spatial dimensions have been normalized with respect
to the PSF width, the PSF width is $1$ in the dimensionless
unit, and the assumption can be expressed as
\begin{align}
\Delta \ll 1,
\end{align}
which defines the subdiffraction regime.  The parameters are then
bounded by 
\begin{align}
|\theta_{\bs\mu}| \le \bk{\frac{\Delta}{2}}^{|\bs\mu|_1},
  \label{small_theta}
\end{align}
and the image is so blurred that it resembles the TEM$_{00}$ mode
rather than the object, viz.,
$f(\bs r|\theta) = \abs{\phi_{00}(\bs r)}^2[1 + O(\Delta)]$.  Writing
the denominator in Eq.~(\ref{Jdirect2}) as $1 + O(\Delta)$ and
applying the orthogonality of Hermite polynomials
\cite{NIST:DLMF,Olver:2010:NHMF}, I obtain
\begin{align}
J_{\bs\mu\bs\nu}^{(\textrm{direct})}
&= \frac{N}{\bs\mu!}\Bk{\delta_{\bs\mu\bs\nu}+O(\Delta)},
\label{Jdirect3}
\\
\textrm{CRB}_{\bs\mu\bs\mu}^{(\textrm{direct})}
&= \frac{\bs\mu!}{N}\Bk{1+O(\Delta)}.
\label{CRBdirect}
\end{align}
This is a significant result in its own right, as it establishes a
fundamental limit to superresolution algorithms for shot-noise-limited
direct imaging \cite{walker93,castro12,candes14,schiebinger},
generalizing the earlier results for two sources
\cite{bettens,vanaert,ram} and establishing that, at least for a
Gaussian PSF, the moments are a natural, approximately orthogonal
\cite{cox87} set of parameters for subdiffraction objects.

\subsection{SPADE}
To investigate the performance of SPADE for moment estimation, note
that, in the subdiffraction regime, Eq.~(\ref{Theta}) can be expressed
as 
\begin{align}
\Theta_{\bs\mu} = \theta_{\bs\mu} + O\bk{\Delta^{|\bs\mu|_1+2}},
\end{align}
where $O(\Delta^{|\bs\mu|_1+2})$ is a linear combination of moments
that are at least two orders above $\bs\mu$ and therefore much smaller
than $\theta_{\bs\mu}$. Approximating $\Theta_{\bs\mu}$ with
$\theta_{\bs\mu}$ greatly simplifies the analysis below;
Appendix~\ref{app_nuisance} contains a more detailed justification of
this approximation.  For the TEM scheme, taking
$\Theta_{2\bs q} = \theta_{2\bs q}$ in Eq.~(\ref{pTEM}) makes the
information matrix diagonal, with the nonzero elements given by
\begin{align}
J_{\bs\mu\bs\mu}^{(\textrm{TEM})} &=
\frac{N^{(\textrm{TEM})}C(\bs q,\bs q)}{\theta_{2\bs q}},
\quad
\bs\mu = 2\bs q,
\label{JTEM}
\end{align}
where $N^{(\textrm{TEM})}$ is the average photon number available to
the TEM scheme. The relevant CRB components are hence
\begin{align}
\textrm{CRB}_{\bs\mu\bs\mu}^{(\textrm{TEM})} &=
\frac{\theta_{2\bs q}}{N^{(\textrm{TEM})}C(\bs q,\bs q)},
\quad
\bs\mu = 2\bs q.
\label{CRBTEM}
\end{align}
Defining the photon count of the $\bs q$th channel as $m_{\bs q}$ with
expected value $N^{(\textrm{TEM})} p^{(\textrm{TEM})}(\bs q|\theta)$, it is
straightforward to show that the estimator
\begin{align}
\check\theta_{2\bs q} = \frac{m_{\bs q}}{N^{(\textrm{TEM})} C(\bs q,\bs q)}
\label{est_TEM}
\end{align}
is unbiased and achieves the error given by Eq.~(\ref{CRBTEM}) under
the assumption $\Theta_{2\bs q} = \theta_{2\bs q}$.

A precision enhancement factor can be defined as the ratio of
Eq.~(\ref{CRBdirect}) to Eq.~(\ref{CRBTEM}), viz.,
\begin{align}
\frac{\textrm{CRB}_{\bs\mu\bs\mu}^{(\textrm{direct})}}
{\textrm{CRB}_{\bs\mu\bs\mu}^{(\textrm{TEM})}}
&\approx \frac{N^{(\textrm{TEM})}}{N}
\frac{\bs\mu!}{2^{|\bs\mu|_1}(\bs\mu/2)!\theta_{\bs\mu}}.
\label{TEM_enhance}
\end{align}
Apart from a factor $N^{(\textrm{TEM})}/N$ determined by the different
photon numbers detectable in each method, the important point is that
the factor scales inversely with
$\theta_{\bs\mu} = O(\Delta^{|\bs\mu|_1})$, so the enhancement is
enormous in the $\Delta \ll 1$ subdiffraction regime.  The prefactor
in Eq.~(\ref{TEM_enhance}) also increases with increasing $\bs\mu$.

To investigate the errors in estimating the other moments via the iTEM
schemes, assume $\Theta_{\bs q+\bs q'} = \theta_{\bs q+\bs q'}$ in
Eqs.~(\ref{piTEM}).  The dependence of Eqs.~(\ref{piTEM}) on
$\theta_{\bs q+\bs q'}$ is the main interest, while I treat
$\beta(\bs q,\bs q')$ as an unknown nuisance parameter \cite{bell};
the TEM scheme can offer additional information about
$\beta(\bs q,\bs q')$ via $\theta_{2\bs q}$ and $\theta_{2\bs q'}$ but
it is insignificant and neglected here to simplify the analysis. The
information matrix with respect to
$\{\theta_{\bs q+\bs q'},\beta(\bs q,\bs q')\}$ is block-diagonal and
consists of a series of two-by-two matrices, each of which can be
determined from Eqs.~(\ref{piTEM}) for two parameters
$(\theta_{\bs q+\bs q'},\beta(\bs q,\bs q'))$ and is given by
\begin{align}
J^{(\bs q,\bs q')}
&= \frac{2N^{(\textrm{iTEM})}}{\beta^2(\bs q,\bs q')
-C^2(\bs q,\bs q')\theta_{\bs q+\bs q'}^2}
\bk{\begin{array}{cc}\beta(\bs q,\bs q') C^2(\bs q,\bs q')
& -C^2(\bs q,\bs q')\theta_{\bs q+\bs q'}\\
-C^2(\bs q,\bs q')\theta_{\bs q+\bs q'} & \beta(\bs q,\bs q')\end{array}},
\label{submatrix}
\end{align}
where $N^{(\textrm{iTEM})}$ is the average photon number available to
the iTEM scheme. The CRB component with respect to
$\theta_{\bs q+\bs q'}$ is hence obtained by taking the inverse of
Eq.~(\ref{submatrix}) and extracting the relevant term; the result is
\begin{align}
\textrm{CRB}_{\bs\mu\bs\mu}^{(\textrm{iTEM})} &=
\frac{\beta(\bs q,\bs q')}{2N^{(\textrm{iTEM})} C^2(\bs q,\bs q')},
&
\bs\mu = \bs q + \bs q'.
\label{CRBiTEM}
\end{align}
Defining the two photon counts of the $(\bs q,\bs q')$ iTEM channels
as $m_+^{(\bs q,\bs q')}$ and $m_-^{(\bs q,\bs q')}$ with expected
values $N^{(\textrm{iTEM})} p^{(\bs q,\bs q')}(+|\theta)$ and
$N^{(\textrm{iTEM})} p^{(\bs q,\bs q')}(-|\theta)$, respectively, it
can be shown that the estimator
\begin{align}
\check\theta_{\bs q+\bs q'} &= \frac{m_+^{(\bs q,\bs q')}-m_-^{(\bs q,\bs q')}}
{2N^{(\textrm{iTEM})} C(\bs q,\bs q')}
\label{est_iTEM}
\end{align}
is unbiased and achieves the error given by Eq.~(\ref{CRBiTEM}) under
the assumption $\Theta_{\bs q+\bs q'} = \theta_{\bs q+\bs q'}$.  The
iTEM schemes can also offer information about $\theta_{2\bs q}$ and
$\theta_{2\bs q'}$ via the background parameter $\beta(\bs q,\bs q')$,
but the additional information is inconsequential and neglected here.

An enhancement factor can again be expressed as
\begin{align}
\frac{\textrm{CRB}_{\bs\mu\bs\mu}^{(\textrm{direct})}}
{\textrm{CRB}_{\bs\mu\bs\mu}^{(\textrm{iTEM})}}
&\approx \frac{N^{(\textrm{iTEM})}}{N}
\frac{\bs\mu!}{2^{2|\bs\mu|_1-1}\bs q! (\bs\mu-\bs q)!\beta(\bs q,\bs\mu-\bs q)}.
\label{iTEM_enhance}
\end{align}
With the background parameter $\beta(\bs q,\bs\mu-\bs q)$ on the order
of $\Delta^{\textrm{min}[|2\bs q|_1,|2(\bs\mu-\bs q)|_1]}$, both
$1/\beta$ and the $\bs\mu!/[\bs q!(\bs\mu-\bs q)!]$ coefficient can be
maximized by choosing $\bs q$ to be as close to $\bs\mu/2$ as
possible. This justifies the pairing of neighboring modes in the iTEM
schemes listed in Fig.~\ref{iTEM} and Table~\ref{projections}. With
iTEM1, iTEM2, iTEM4, and iTEM5, $|\bs\mu|_1$ is odd, and
\begin{align}
\beta = O\bk{\Delta^{|\bs\mu|_1-1}}.
\end{align}
With iTEM3 and iTEM6, $|\bs\mu|_1$ is even, and
\begin{align}
\beta = O\bk{\Delta^{|\bs\mu|_1}}.
\end{align}
The enhancements, being inversely proportional to $\beta$, can again
be substantial for higher moments. The only exception is the
estimation of the first moments $\theta_{10}$ and $\theta_{01}$, for
which the right-hand side of Eq.~(\ref{iTEM_enhance}) becomes
$N^{(\textrm{iTEM})}/N$ and the iTEM schemes offer no advantage.

These results can be compared with Refs.~\cite{tnl,ant} for the
special case of two equally bright point sources. If the origin of the
image plane is aligned with their centroid and their separation along
the $X$ direction is $d$, $\theta_{20} = d^2/4$, and a
reparameterization leads to a transformed Fisher information
$\mathcal J^{(\textrm{direct})}(d) \approx N d^2/8$ and
$\mathcal J^{(\textrm{TEM})}(d) \approx N/4$ for the estimation of
$d$, in accordance with the results in Refs.~\cite{tnl,ant} to the
leading order of $d$.
The experiments reported in Refs.~\cite{yang16,tham16,paur16}
serve as demonstrations of the proposed scheme in this special case.

\subsection{Elementary explanation}
\label{sec_explain}
The enhancements offered by SPADE can be understood by considering the
signal-to-noise ratio (SNR) of a measurement with Poisson statistics.
Suppose for simplicity that the mean count of an output can be written
as $N p(j|\theta) = A \theta + B$, which consists of a signal
component $A \theta$ and a background $B$. The variance is
$A\theta + B$, so the SNR can be expressed as
$(A\theta)^2/(A\theta + B)$. To maximize it, the background $B$ should
be minimized to reduce the variance. For direct imaging, the
background according to Eq.~(\ref{lambda_linear}) is dominated by the
TEM$_{00}$ mode, whereas each output of SPADE is able to filter out
that mode as well as other irrelevant low-order modes to minimize the
background without compromising the signal.  To wit, Eq.~(\ref{pTEM})
for TEM measurements has no background, while Eqs.~(\ref{piTEM}) for
iTEM also have low backgrounds in the subdiffraction regime. The
Fisher information given by Eq.~(\ref{fisher}) is simply a more
rigorous statistical tool that formalizes the SNR concept and provides
error bounds; reducing the background likewise improves the
information by reducing the denominator in Eq.~(\ref{fisher}).

In this respect, the proposed scheme seems to work in a similar way to
nulling interferometry for exoplanet detection \cite{labeyrie}. The
nulling was proposed there for the special purpose of blocking the
glare of starlight, however, and there had not been any prior
statistical study of nulling for subdiffraction objects to my
knowledge. The surprise here is that coherent processing in the far
field can vastly improve general incoherent imaging even in the
subdiffraction regime and in the presence of photon shot noise,
without the need to manipulate the sources as in prior superresolution
microscopic methods
\cite{moerner,betzig,hell,kolobov07,taylor2013,localization} or detect
evanescent waves via lossy or unrealistic materials
\cite{pendry00,tsang08}.

\section{Numerical demonstration}
\label{sec_numerical}
\begin{figure}[htbp!]
\centerline{\includegraphics[width=0.6\textwidth]{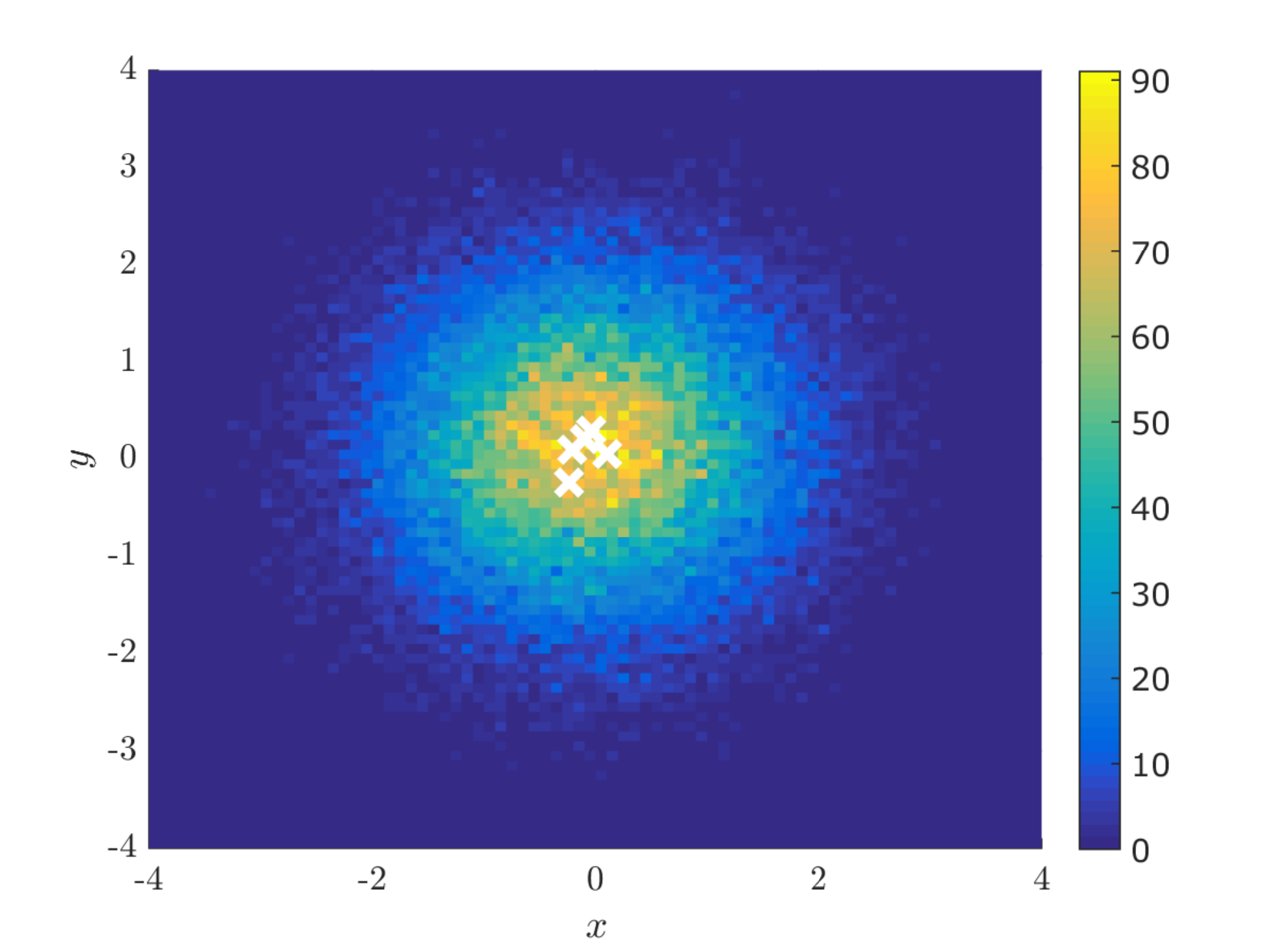}}
\caption{The white crosses denote the 5 randomly generated source
  positions. The background image is a direct image with pixel size
  $dx dy = 0.1\times 0.1$ (normalized with respect to the PSF width)
  and Poisson noise; the average photon number is $N = 5\times 10,000$
  in total.}
\label{direct_image}
\end{figure}

Here I present a numerical study to illustrate the proposal and
confirm the theory.  Assume an object that consists of 5 equally
bright point sources with random positions within the square
$-0.3 \le X \le 0.3$ and $-0.3 \le Y \le 0.3$.  The average photon
number is assumed to be $N = 5\times 10,000$ in
total. Figure~\ref{direct_image} shows an example of the generated
source positions and a direct image with pixel size
$\delta x \delta y = 0.1\times 0.1$ and Poisson noise.  I focus on the
estimation of the first and second moments of the source distribution
given by
$\{\theta_{\bs\mu}; \bs\mu = (1,0), (0,1), (2,0), (0,2),(1,1)\}$.  For
direct imaging, I use the estimator
$\check\theta_{\bs\mu} = \bs\mu!\sum_{j} D_{\bs\mu}(\bs r_j)m(\bs
r_j)/N$, where $m(\bs r_j)$ is the photon count at a pixel positioned
at $\bs r_j$. It can be shown that, in the small-pixel limit, this
estimator is unbiased and approaches the CRB given by
Eq.~(\ref{CRBdirect}) for $\Delta\ll 1$.

\begin{figure}[htbp!]
\centerline{\includegraphics[width=0.6\textwidth]{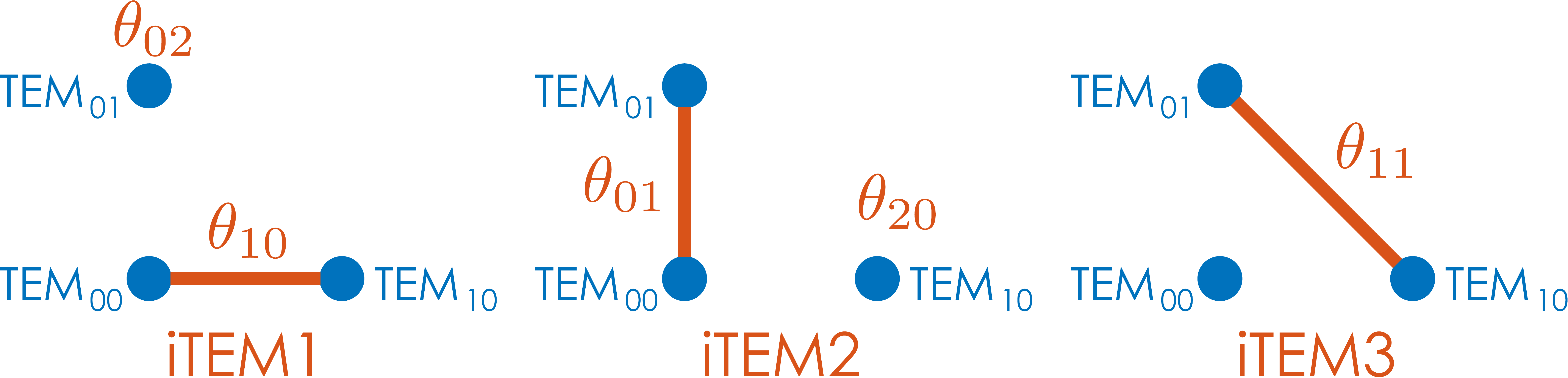}}
\caption{A graphical representation of the iTEM1, iTEM2, and iTEM3
  schemes involving the three TEM modes to be measured. Each line
  denotes an interferometer between two modes, and each unconnected
  dot denotes a TEM mode to be measured.  The modes are also denoted
  by the parameters $\theta_{\bs\mu}$ to which they are sensitive.}
\label{iTEMlow}
\end{figure}

\begin{figure}[htbp!]
\centerline{\includegraphics[width=0.6\textwidth]{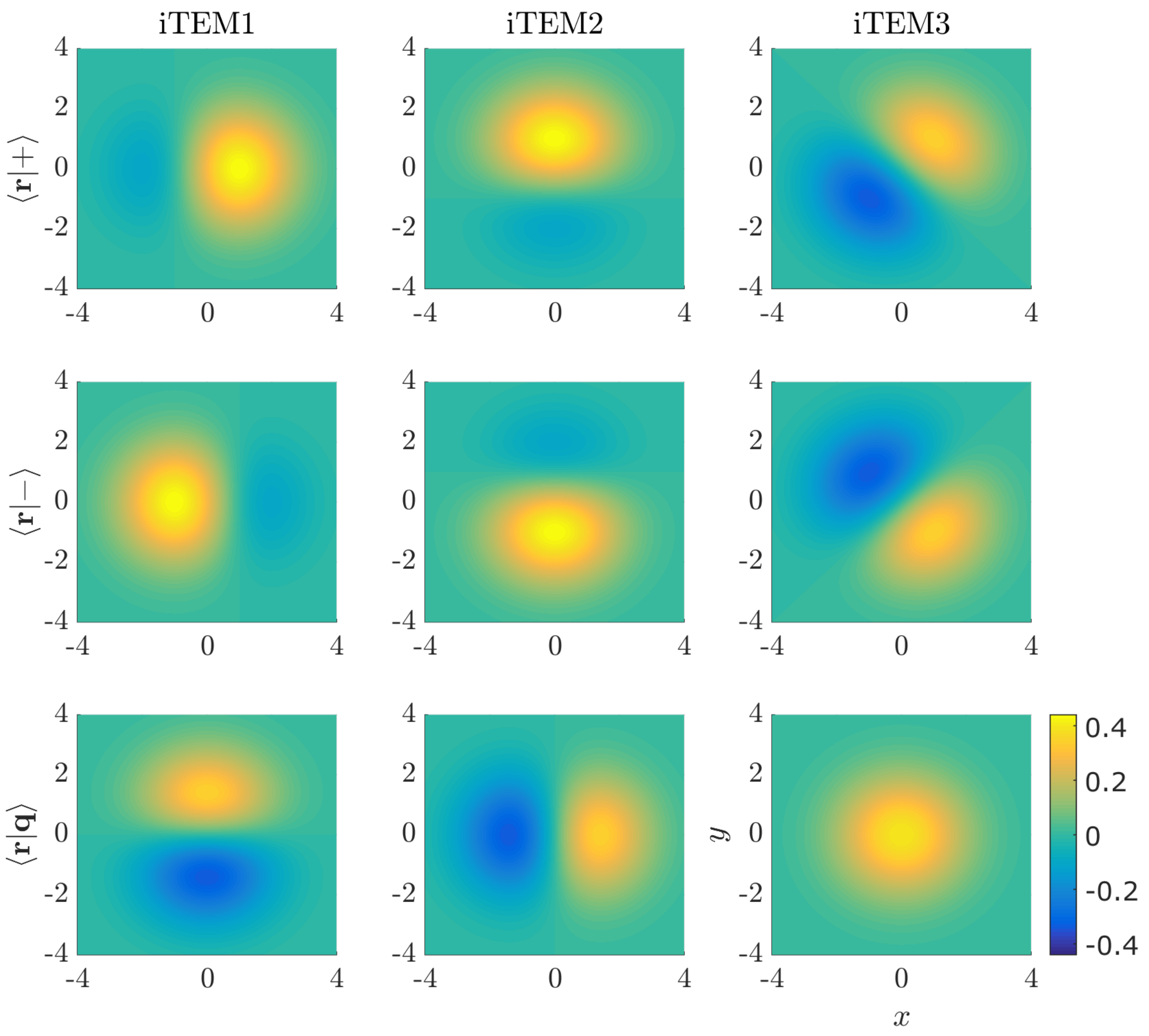}}
\caption{The spatial wave functions $\braket{\bs r|\phi_j}$ for the
  projections listed in Table~\ref{iTEM_table}. $x$ and $y$ are
  image-plane coordinates normalized with respect to the PSF width and
  the color code corresponds to amplitudes of normalized wave
  functions.}
\label{TEMplot}
\end{figure}

\begin{figure}[htbp!]
\centerline{\includegraphics[width=0.6\textwidth]{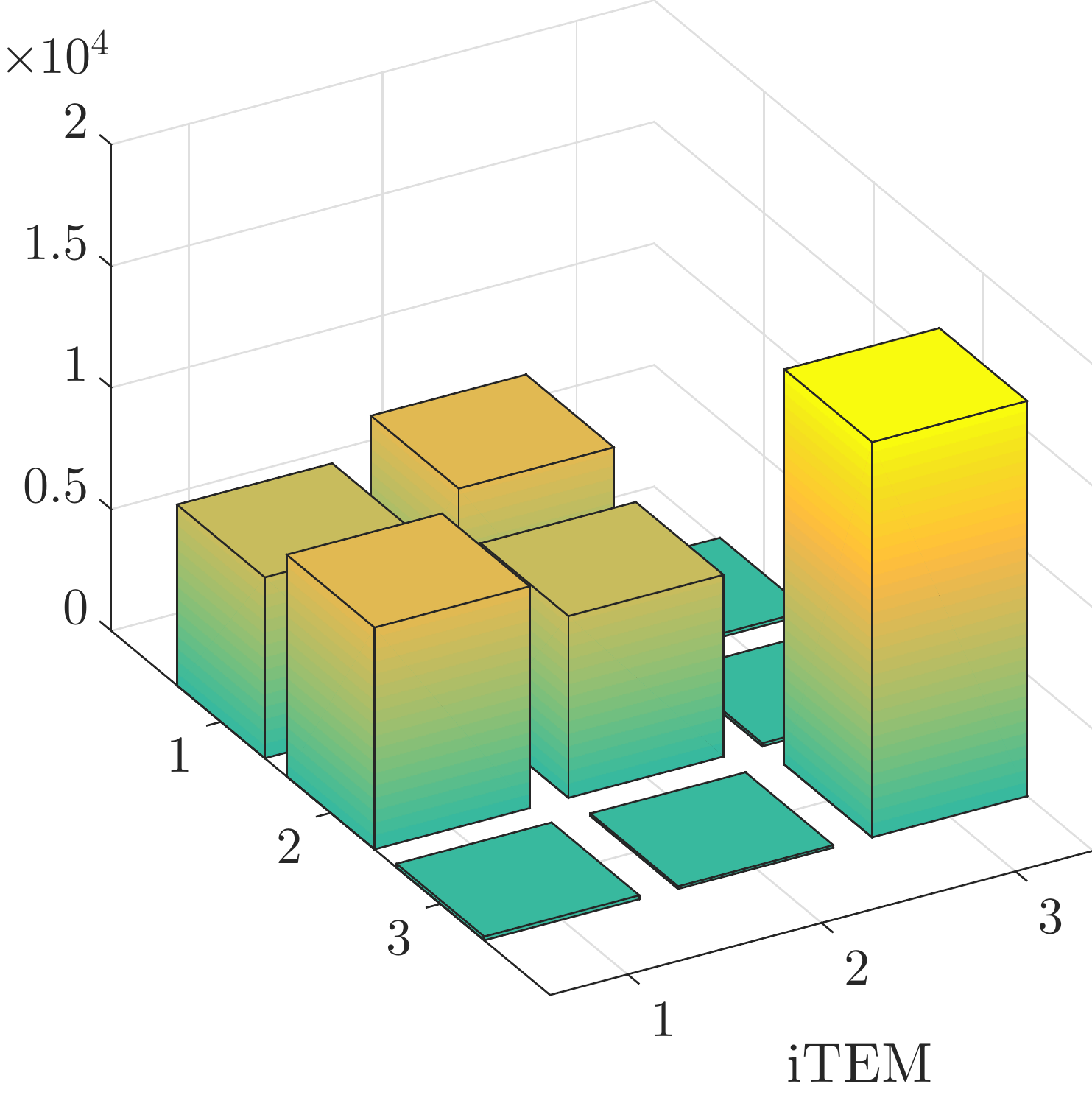}}
\caption{A sample of the simulated photon counts from SPADE. The order
  of the matrix elements follows Table~\ref{iTEM_table} and
  Fig.~\ref{TEMplot}. Note how the counts for the antisymmetric modes
  are much lower as a result of filtering out the lower-order modes.
  As argued in Sec.~\ref{sec_explain}, such dark ports
  enable a higher SNR by reducing the background without compromising
  the signal.  }
\label{iTEM_counts}
\end{figure}

For SPADE, I consider only the TEM$_{00}$, TEM$_{10}$, and TEM$_{01}$
modes, and the photons in all the other modes are discarded.  As
illustrated in Fig.~\ref{iTEMlow}, the iTEM1, iTEM2, and iTEM3 schemes
suffice to estimate the parameters of interest. Table~\ref{iTEM_table}
lists the projections, and Fig.~\ref{TEMplot} plots the spatial wave
functions for the projections.  The light is assumed to be split
equally among the three schemes, leading to 9 outputs;
Fig.~\ref{iTEM_counts} shows a sample of the photon counts drawn from
Poisson statistics. For the estimators, I use Eq.~(\ref{est_TEM}) and
(\ref{est_iTEM}).  Compared with the large number of pixels in direct
imaging, the compressive nature of SPADE for moment estimation is an
additional advantage.

\begin{table}[htbp!]
\centerline{
\begin{tabularx}{0.8\textwidth}{|X|X|X|}
\hline
iTEM1 & iTEM2 & iTEM3\\
\hline
$(\ket{0,0}+\ket{1,0})/\sqrt{2}$
& $(\ket{0,0}+\ket{0,1})/\sqrt{2}$ & 
$(\ket{1,0}+\ket{0,1})/\sqrt{2}$  \\
$(\ket{0,0}-\ket{1,0})/\sqrt{2}$ & 
$(\ket{0,0}-\ket{0,1})/\sqrt{2}$ & 
$(\ket{1,0}-\ket{0,1})/\sqrt{2}$ \\
$\ket{0,1}$ & $\ket{1,0}$ & $\ket{0,0}$\\
\hline
\end{tabularx}
}
\caption{The projections for the SPADE measurement scheme depicted
  in Fig.~\ref{iTEMlow}. $\ket{0,0}$ corresponds to the TEM$_{00}$ mode,
  $\ket{1,0}$ corresponds to the TEM$_{10}$ mode, and $\ket{0,1}$
  corresponds to the TEM$_{01}$ mode.}
\label{iTEM_table}
\end{table}

\begin{figure}[htbp!]
\centerline{\includegraphics[width=0.6\textwidth]{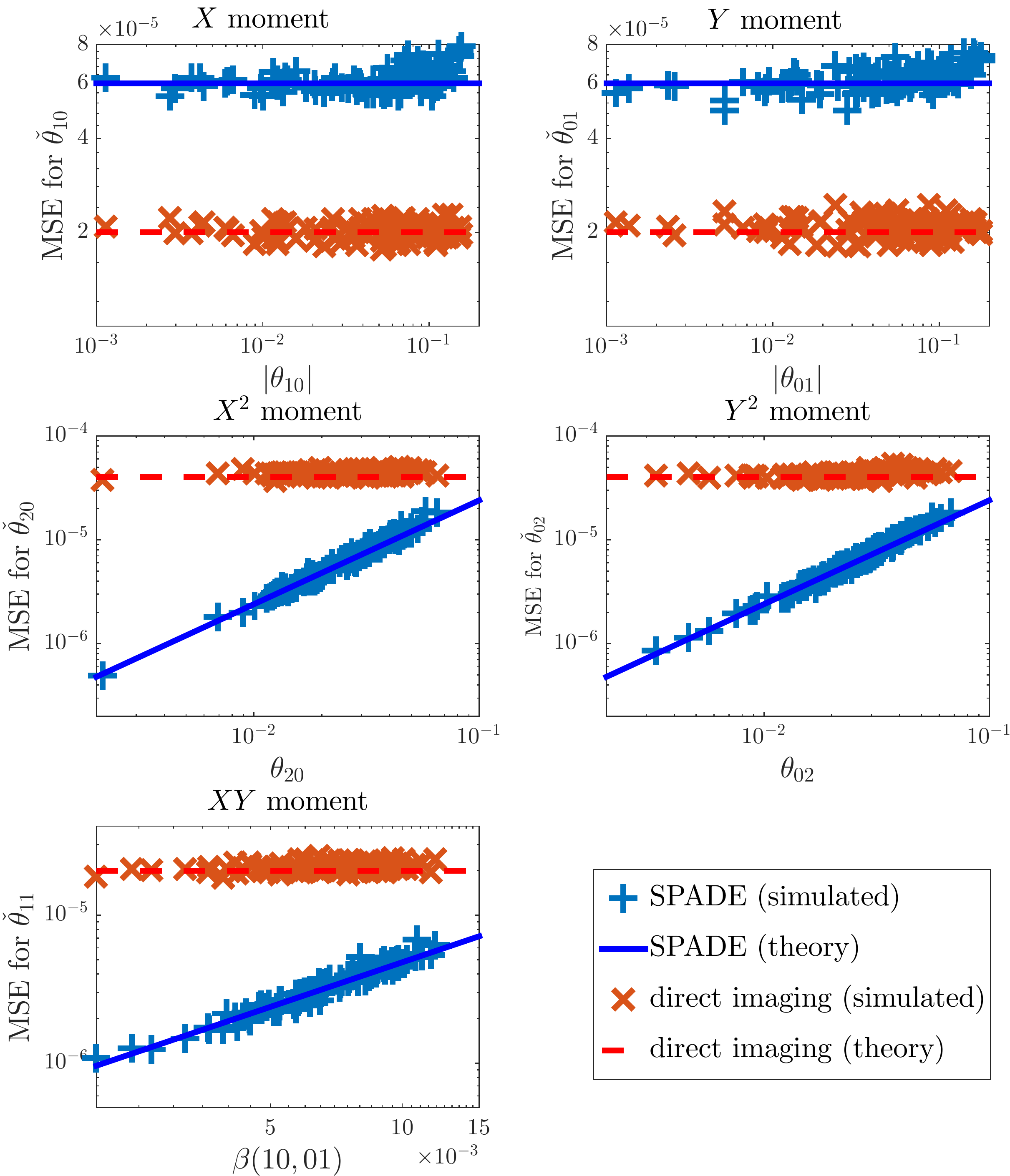}}
\caption{Simulated errors for SPADE and direct imaging versus certain
  parameters of interest in log-log scale. The lines are the CRBs
  given by Eqs.~(\ref{CRBdirect}), (\ref{CRBTEM}), and
  (\ref{CRBiTEM}), assuming $\Theta_{\bs\mu} = \theta_{\bs\mu}$ and
  neglecting the $O(\Delta)$ term in Eq.~(\ref{CRBdirect}). Recall
  that all lengths are normalized with respect to the PSF width
  $\sigma$, so, in real units, the first moments $\theta_{10}$ and
  $\theta_{01}$ are in units of $\sigma$, their MSEs are in units of
  $\sigma^2$, the second moments $\theta_{20}$, $\theta_{02}$,
  $\theta_{11}$, $\beta(10,01) = (\theta_{20}+\theta_{02})/8$ are in
  units of $\sigma^2$, and their MSEs are in units of $\sigma^4$.}
\label{errors}
\end{figure}

Figure~\ref{errors} plots the numerically computed mean-square errors
(MSEs) for 100 randomly generated objects versus true parameters in
log-log scale. Each error value for a given object is computed by
averaging the squared difference between the estimator and the true
parameter over 500 samples of Poissonian outputs. For comparison,
Fig.~\ref{errors} also plots the CRBs given by 
Eqs.~(\ref{CRBdirect}), (\ref{CRBTEM}), and (\ref{CRBiTEM}), assuming
$\Theta_{\bs\mu} = \theta_{\bs\mu}$ and neglecting the $O(\Delta)$ term
in Eq.~(\ref{CRBdirect}).  A few observations can be made:
\begin{enumerate}
\item As shown by the plots in the first row of Fig.~\ref{errors},
  SPADE is 3 times worse than direct imaging at estimating the first
  moments. This is because SPADE uses only 1/3 of the available
  photons to estimate each first moment.

\item The theory suggests that the advantage of SPADE starts with the
  second moments, and indeed the other plots show that SPADE is
  substantially more precise at estimating them, even though SPADE
  uses only a fraction of the available photons to estimate each
  moment. This enhancement is a generalization of the recent results
  on two sources
  \cite{tnl,ant,sliver,nair_tsang16,tnl2,tsang16,lupo,rehacek16,tang16,yang16,tham16,paur16}.

\item The errors are all remarkably tight to the CRBs, despite the
  simplicity of the estimators and the approximations in the bounds.
  In particular, the excellent performance of the SPADE estimator in
  the subdiffraction regime justifies its assumption of
  $\Theta_{\bs\mu}=\theta_{\bs\mu}$.
\end{enumerate}

\section{Quantum limits}
\label{sec_limits}
In the diffraction-\emph{un}limited regime, it is not difficult to
prove that direct imaging achieves the highest Fisher information
allowed by quantum mechanics. To be precise, that regime can be
defined as one in which the PSF is so sharp relative to the source
distribution that
$\{\ket{\psi_{\bs R}}; \bs R \in \operatorname{supp}(F)\}$ can be
approximated as the orthogonal position basis $\{\ket{\bs
  r}\}$. $\rho_1$ becomes diagonal in that basis, and the quantum
Fisher information \cite{helstrom,holevo11,hayashi05,hayashi,tnl,tnl2}
is equal to the direct-imaging information given by
Eq.~(\ref{Jdirect}).  The physics in the opposite subdiffraction
regime is entirely different, however, as diffraction causes
$\{\ket{\psi_{\bs R}}\}$ to have significant overlaps with one
another, and more judicious measurements can better deal with the
resulting indistinguishability, as demonstrated in
Secs.~\ref{sec_stat} and \ref{sec_numerical}.

I now prove that SPADE is in fact near-quantum-optimal in estimating
location and scale parameters of a source distribution in the
subdiffraction regime. Suppose that the distribution has the form
\begin{align}
 F(\bs R|\theta) &=  F(\bs R(\bs\xi|\theta)),
\end{align}
such that $\theta$ parameterizes a coordinate transformation
$\bs R = \bs R(\bs\xi|\theta)$, and the transformation leads to a
reference measure $F_0(\bs\xi)$ that is independent of $\theta$.
Taking $\bs R$ as a column vector, I can rewrite Eq.~(\ref{rho1}) as
\begin{align}
\rho_1(\theta) &= \expect_0\bk{
\ket{\psi_{\bs R}}\bra{\psi_{\bs R}}},
\\
\ket{\psi_{\bs R}} &= e^{-i \bs k^\top \bs R(\bs\xi|\theta)}\ket{\psi_{\bs 0}},
\end{align}
where $\expect_0(\cdot) \equiv \int d\bs\xi F_0(\bs\xi)(\cdot)$ and
$\bs k$ is the momentum operator in a column vector.  I can now use
the quantum upper bound on the Fisher information
\cite{hayashi05,hayashi,tnl} and the convexity of the quantum Fisher
information \cite{fujiwara01,ng16} to prove that the Fisher
information for any measurement is bounded as
\begin{align}
J(\theta) &\le K(\rho_1(\theta)) \le \tilde K(\theta) \equiv N \expect_0
\Bk{K\bk{\ket{\psi_{\bs R}}\bra{\psi_{\bs R}}}},
\end{align}
where $K$ is the quantum Fisher information proposed by Helstrom
\cite{helstrom}.  For the pure state, $K$ can be computed analytically
to give
\begin{align}
K\bk{\ket{\psi_{\bs R}}\bra{\psi_{\bs R}}} &= 
4\parti{\bs R^\top(\bs\xi|\theta)}{\theta_\mu}
\bra{\psi_{\bs 0}}\Delta\bs k\Delta\bs k^\top \ket{\psi_{\bs 0}}
\parti{\bs R(\bs\xi|\theta)}{\theta_\nu},
\\
\Delta\bs k &\equiv \bs k-\bra{\psi_0}\bs k\ket{\psi_0},
\end{align}
leading to
\begin{align}
\tilde K_{\mu\nu} &= N
\expect_0\Bk{\parti{\bs R^\top(\bs\xi|\theta)}{\theta_\mu}
\parti{\bs R(\bs\xi|\theta)}{\theta_\nu}}
\label{Jtilde}
\end{align}
for the gaussian PSF. For example, a location parameter can be
expressed as $\bs R = \bs\xi -(\theta,0)$.  Equation~(\ref{Jtilde})
then gives
\begin{align}
\tilde K &= N.
\end{align}
This can be attained by either direct imaging or iTEM1 in the
subdiffraction regime. 

The advantage of SPADE starts with the second moments, which are
particularly relevant to scale estimation.  Let
$\bs R = \theta\bs\xi$, which results in
\begin{align}
\tilde K &= N \expect_0\bk{\bs\xi^\top \bs\xi}.
\label{Jscale}
\end{align}
For the TEM measurement, on the other hand,
\begin{align}
p^{(\textrm{TEM})}(\bs q|\theta) &= 
\expect_0\bk{e^{-Q_X} \frac{Q_X^{q_x}}{q_x!}
e^{-Q_Y} \frac{Q_Y^{q_y}}{q_y!}},
\\
Q_X &\equiv \frac{\theta^2\xi_X^2}{4},
\quad
Q_Y \equiv \frac{\theta^2\xi_Y^2}{4}.
\end{align}
Defining
\begin{align}
\bar{\bs q} &\equiv \sum_{\bs q} \bs q 
p^{(\textrm{TEM})}(\bs q|\theta) = \frac{\theta^2}{4}
\expect_0\bk{\begin{array}{c}\xi_X^2\\ \xi_Y^2\end{array}},
\\
\bs V &\equiv \sum_{\bs q} 
\bk{\bs q-\bar{\bs q}} 
\bk{\bs q-\bar{\bs q}}^\top p^{(\textrm{TEM})}(\bs q|\theta)
\\
&=
\frac{\theta^2}{4}
\expect_0\bk{\begin{array}{cc}\xi_X^2 & 0 \\ 0 & \xi_Y^2\end{array}}
+O(\theta^4),
\label{V}
\end{align}
and using the lower bound by Stein \textit{et al.}~\cite{stein14}, I
obtain
\begin{align}
J^{(\textrm{TEM})} &\ge N
\parti{\bar{\bs q}^\top}{\theta} \bs V^{-1}\parti{\bar{\bs q}}{\theta} 
\to N\expect_0\bk{\bs\xi^\top\bs\xi},
\end{align}
which approaches the quantum limit given by Eq.~(\ref{Jscale}) for
$\theta \to 0$.  The argument can be made more precise if the form of
$F_0(\bs\xi)$ is known, as the extended convexity of the quantum
Fisher information can be used to obtain a tighter upper bound
\cite{alipour,ng16}, while the $O(\theta^4)$ term in Eq.~(\ref{V}) can
be computed to obtain an explicit lower bound for any $\theta$. 

\section{Discussion}
\label{sec_discuss}
Though promising, the giant precision enhancements offered by SPADE do
not imply unlimited imaging resolution for finite photon numbers.  The
higher moments are still more difficult to estimate even with SPADE in
terms of the fractional error, which is
$\sim \textrm{CRB}_{\bs\mu\bs\mu}/\theta_{\bs\mu}^2 =
1/O(N\Delta^{|\bs\mu|_1})$ for even $|\bs\mu|_1$ and
$1/O(N\Delta^{|\bs\mu|_1+1})$ for odd $|\bs\mu|_1$, meaning that more
photons are needed to attain a desirable fractional error for
higher-order moments. Intuitively, this is because of the inherent
inefficiency of subdiffraction objects to couple to higher-order
modes, and the need to accumulate enough photons in those modes to
achieve an acceptable SNR. A related issue is the reconstruction of
the full source distribution, which requires all moments in principle.
A finite number of moments cannot determine the distribution uniquely
by themselves \cite{bertero89}, although a wide range of
regularization methods, such as maximum entropy and basis pursuit, are
available for more specific scenarios
\cite{bertero89,gull84,milanfar96,castro12,schiebinger}.

Despite these limitations, the fact remains that direct imaging is an
even poorer choice of measurement for subdiffraction objects and SPADE
can extract much more information, simply from the far field. For
example, the size and shape of a star, a planetary system, a galaxy,
or a fluorophore cluster that is poorly resolved under direct imaging
can be identified much more accurately through the estimation of the
second or higher moments by SPADE. Alternatively, SPADE can be used to
reach a desirable precision with far fewer photons or a much smaller
aperture, enhancing the speed or reducing the size of the imaging
system for the more special purposes.  In view of the statistical
analysis in Refs.~\cite{sliver,ant,nair_tsang16}, the image-inversion
interferometers proposed in
Refs.~\cite{wicker07,wicker09,weigel,sliver,ant,nair_tsang16} are
expected to be similarly useful for estimating the second moments. For
larger objects, scanning in the manner of confocal microscopy
\cite{pawley} or adaptive alignment \cite{tnl} should be helpful.

Many open problems remain; chief among them are the incorporation of
prior information, generalizations for non-Gaussian PSFs, the
derivation of more general quantum limits, the possibility of even
better measurements, and experimental implementations.  The quantum
optimality of SPADE for general imaging is in particular an
interesting question. These daunting problems may be attacked by more
advanced methods in quantum metrology
\cite{helstrom,holevo11,hayashi05,hayashi,kahn09,qbzzb,qwwb,gagatsos16},
quantum state tomography \cite{jezek,lvovsky09,gross10}, compressed
sensing \cite{castro12,candes14,schiebinger,gross10}, and photonics
design \cite{armstrong,miller13,li14}.

\section*{Acknowledgments}
Inspiring discussions with Ranjith Nair, Xiao-Ming Lu, Shan Zheng Ang,
Shilin Ng, Laura Waller's group, Geoff Schiebinger, Ben Recht, and
Alex Lvovsky are gratefully acknowledged.  This work is supported by
the Singapore National Research Foundation under NRF Grant
No.~NRF-NRFF2011-07 and the Singapore Ministry of Education Academic
Research Fund Tier 1 Project R-263-000-C06-112.

\appendix

\section{\label{app_nuisance}Nuisance parameters}
Instead of assuming $\Theta_{\bs\mu} = \theta_{\bs\mu}$ as in
Sec.~\ref{sec_stat}, I consider here the exact relation
given by Eq.~(\ref{Theta}), which can be expressed as
\begin{align}
\Theta_{\bs\mu} &= \theta_{\bs\mu} - \frac{\theta_{\bs\mu +(2,0)}}{4}
- \frac{\theta_{\bs\mu +(0,2)}}{4} - \dots
\label{Theta_series}
\end{align}
For the TEM scheme, this implies that each $\bs q$th channel contains
information about not only $\theta_{2\bs q}$ but also the higher-order
moments. If I assume that each $\bs q$th channel is used to estimate
only $\theta_{2\bs q}$, however, then the higher-order moments act
only as nuisance parameters \cite{bell} to the estimation of
$\theta_{2\bs q}$.  This is a conservative assumption, as the
data-processing inequality \cite{zamir,hayashi} implies that
neglecting outputs can only reduce the information, but the assumption
also means that I do not need to consider any channel with order lower
than $\bs q$ to compute the CRB with respect to $\theta_{2\bs q}$,
simplifying the analysis below.

Given the above assumption, I can compute the information matrix with
respect to the parameters
$(\theta_{2\bs q}, \theta_{2\bs q+(2,0)},\theta_{2\bs q+(0,2)},\dots)$
by considering only the $\bs q$th and higher-order channels; the
result is
\begin{align}
J^{(\textrm{TEM})} &= \bk{\begin{array}{c|cc}
J_{2\bs q,2\bs q}^{(\textrm{TEM})} & J_{2\bs q, 2\bs q+(2,0)}^{(\textrm{TEM})} & \dots\\
\hline
J_{2\bs q+(2,0),2\bs q}^{(\textrm{TEM})} & J_{2\bs q+(2,0),2\bs q+(2,0)}^{(\textrm{TEM})}
& \dots\\
\vdots & \vdots & \ddots
\end{array}}
\equiv N^{(\textrm{TEM})}
\bk{\begin{array}{c|c}
\alpha & \eta^\top\\
\hline
\eta & \gamma
\end{array}},
\end{align}
where $\alpha = C(\bs q,\bs q)/\Theta_{2\bs q}$ and
$\eta = O(\Delta^{-2|\bs q|_1})$ are determined only by the $\bs q$th
channel, while $\gamma = O(\Delta^{-(2|\bs q|_1+2)})$ is mainly
determined by the higher-order channels.  The CRB with respect to
$\theta_{2\bs q}$ becomes \cite{bell}
\begin{align}
\textrm{CRB}^{(\textrm{TEM})}_{2\bs q,2\bs q} = \frac{1}{N^{(\textrm{TEM})}}
\bk{\alpha - \eta^\top\gamma^{-1}\eta}^{-1}.
\label{nuisance} 
\end{align}
The key here is that $\eta\gamma^{-1}\eta^\top = O(\Delta^{-2|\bs q|_1+2})$
is smaller than $\alpha = O(\Delta^{-2|\bs q|_1})$ by two orders
of $\Delta$, so 
\begin{align}
  \textrm{CRB}^{(\textrm{TEM})}_{2\bs q,2\bs q}
  &= \frac{1}{N^{(\textrm{TEM})}\alpha}\Bk{1+O\bk{\Delta^2}}
  \\
  &= \frac{\Theta_{2\bs q}}{N^{(\textrm{TEM})}
    C(\bs q,\bs q)} \Bk{1+O\bk{\Delta^2}}
  \\
  &=
    \frac{\theta_{2\bs q}}{N^{(\textrm{TEM})}
    C(\bs q,\bs q)} \Bk{1+O\bk{\Delta^2}},
\end{align}
which is consistent with Eq.~(\ref{CRBTEM}).

An intuitive way of understanding this result
is to rewrite Eq.~(\ref{Theta_series}) as
\begin{align}
\theta_{\bs\mu} &= \Theta_{\bs\mu} +  \frac{\theta_{\bs\mu +(2,0)}}{4}
+ \frac{\theta_{\bs\mu +(0,2)}}{4} + \dots,
\end{align}
which implies that the total error in $\theta_{\bs\mu}$ consists of
the error in $\Theta_{\bs\mu}$ as well as the errors in the
higher-order moments. The higher-order moments can be estimated much
more accurately via the higher-order channels, so the effect of their
uncertainties on the estimation of $\theta_{\bs\mu}$ is negligible. A
similar exercise can be done for the iTEM schemes, with similar
results.

In practice, such a careful treatment of the nuisance parameters is
unlikely to be necessary in the subdiffraction regime, as the
numerical analysis in Sec.~\ref{sec_numerical} shows that excellent
results can be obtained simply by taking
$\Theta_{\bs\mu} = \theta_{\bs\mu}$ without any correction.


\bibliography{diameter_estimation_paper6}

\end{document}